\documentclass[aps,pre,twocolumn,showpacs,superscriptaddress]{revtex4-1}

\usepackage[final]{graphicx}
\usepackage{amsmath}
\usepackage{verbatim}
\usepackage{color}
\usepackage{ulem}

\usepackage{threeparttable}

\usepackage{multirow}
\usepackage[colorlinks,
linkcolor=blue,
citecolor=blue,
urlcolor=blue]{hyperref}

\begin{document}

\preprint{}

\title{Stimulated Brillouin scattering behaviors in different species ignition hohlraum plasmas in high-temperature and high-density region}

\author{Q. S. Feng} 
\affiliation{HEDPS, Center for
	Applied Physics and Technology, Peking University, Beijing 100871, China}

\author{C. Y. Zheng} \email{zheng\_chunyang@iapcm.ac.cn}

\affiliation{HEDPS, Center for
	Applied Physics and Technology, Peking University, Beijing 100871, China}
\affiliation{Institute of Applied Physics and Computational
	Mathematics, Beijing, 100094, China}
\affiliation{Collaborative Innovation Center of IFSA (CICIFSA) , Shanghai Jiao Tong University, Shanghai, 200240, China}

\author{Z. J. Liu} 
\affiliation{HEDPS, Center for
	Applied Physics and Technology, Peking University, Beijing 100871, China}
\affiliation{Institute of Applied Physics and Computational
	Mathematics, Beijing, 100094, China}

\author{Q. Wang}
\affiliation{HEDPS, Center for
	Applied Physics and Technology, Peking University, Beijing 100871, China}

\author{C. Z. Xiao}
\affiliation{School of Physics and Electronics, Hunan University, Changsha 410082, China}

\author{L. H. Cao} 
\affiliation{HEDPS, Center for
	Applied Physics and Technology, Peking University, Beijing 100871, China}
\affiliation{Institute of Applied Physics and Computational
	Mathematics, Beijing, 100094, China}
\affiliation{Collaborative Innovation Center of IFSA (CICIFSA) , Shanghai Jiao Tong University, Shanghai, 200240, China}

\author{X. T. He} \email{xthe@iapcm.ac.cn}
\affiliation{HEDPS, Center for
	Applied Physics and Technology, Peking University, Beijing 100871, China}
\affiliation{Institute of Applied Physics and Computational
	Mathematics, Beijing, 100094, China}
\affiliation{Collaborative Innovation Center of IFSA (CICIFSA) , Shanghai Jiao Tong University, Shanghai, 200240, China}


\date{\today}

\begin{abstract}
The presence of multiple ion species can add additional branches to the IAW dispersion relation and change the Landau damping significantly. Different IAW modes excited by stimulated Brillouin scattering (SBS) and different SBS behaviors in several typical ignition hohlraum plasmas in the high-temperature and high-density region have been researched by Vlasov-Maxwell simulation. The slow mode in HeH or CH plasmas is the least damped mode and will be excited in SBS, while the fast mode in AuB plasmas is the least damped mode and will be excited in SBS. Due to strong Landau damping, the SBS in H or HeH plasmas is strong convective instability, while the SBS in AuB plasmas is absolute instability due to the weak Landau damping. However, although the SBS in CH plasmas is weak convective instability in the linear theory, the SBS will transform into absolute instability due to decreasing linear Landau damping by particles trapping. These results give a detail research of the IAW modes excitation and the properties of SBS in different species plasmas, thus providing the possibility of controlling SBS by increasing the linear Landau damping of the IAW by changing ion species.
	
\end{abstract}


\maketitle

\section{\label{Sec: Introduction}Introduction}
Backward stimulated Brillouin scattering (SBS) is a three-wave interaction process where an incident electromagnetic wave (EMW) decays into a backscattered EMW and a forward propagating ion-acoustic wave (IAW). Backward SBS leads to a great energy loss of the incident laser and is detrimental in inertial confinement fusion (ICF) \cite{He_2016POP,Glenzer_2010Science,Glenzer_2007Nature}, including indirect-drive ignition \cite{Lindl_1995POP,Lindl_2004POP}, direct-drive ignition \cite{Nuckolls_1972Nature,Bodner_1998POP}, shock ignition \cite{Betti_2007PRL,Hao_2016POP}, and hybrid-drive ignition \cite{He_2016POP} with a spherical hohlraum \cite{LanKe_2017PRE,Lan_2016MRE,Huo_2016MRE,Huo_2016PRL} where indirect-drive and direct-drive processes are combined. Therefore, SBS plays an important role in the successful ignition goal of ICF. 
Multiple ion species are contained in the laser fusion program \cite{Lindl_2004POP,Neumayer_2008PRL,Neumayer_2008POP,Berger_2015PRE}. In the indirect-drive ICF, the inside of hohlraum will be filled with low-Z plasmas, such as H or HeH plasmas, from the initial fill material; low-Z and mid-Z mixing plasmas, such as CH plasmas from the ablated material off the capsule; and high-Z mixed with mid-Z plasmas, such as AuB plasmas, blowoff from the hohlraum wall surfaces \cite{Froula_2006POP,Meezan_2010POP,Froula_2002PRL}. Understanding the IAW modes excitation or growth and the SBS saturation mechanisms in these different plasmas is vitally necessary to predict SBS laser energy losses and to improve the energy coupling into the fusion capsule. The presence of multi-ion species can add additional branches to the IAW dispersion relation and change the total Landau damping significantly, which may provide the possibility of controlling SBS by increasing the linear damping of the IAW \cite{Neumayer_2008POP,Neumayer_2008PRL}. In this paper, the different IAW modes excited in SBS in different species plasmas have been demonstrated, the theoretical results are consistent to the simulations very well. There exists only one group of modes in the single-ion species plasmas, such as H plasmas; while there exist two groups of modes in multi-ion species plasmas called \textquotedblleft the fast mode" and \textquotedblleft the slow mode" \cite{Feng_2016POP,Chapman_2013PRL}. The least damped IAW mode in HeH or CH plasmas is the slow mode, while the least damped mode in AuB plasmas is the fast mode. And the kinetic theory and Vlasov simulation unambiguously clarify that the least damped mode will be excited and of the largest growth rate in SBS.  

Many methods for the saturation of SBS have been proposed, including increasing linear Landau damping by kinetic ion heating \cite{Rambo_1997PRL, Pawley_1982PRL}, the creation of cavitons in plasmas \cite{Liu_2009POP_1, Weber_2005PRL, Weber_2005POP,Weber_2005POP_1}, frequency detuning due to particles trapping \cite{Froula_2002PRL,Giacone_1998POP,Vu_2001PRL,Albright_2016POP}, coupling with higher harmonics \cite{Bruce_1997POP, Rozmus_1992POP}, and so on. However, the accurate modes analyses of IAWs exited by SBS and the features of SBS in different species plasmas call for deep research. In this paper, for different species ignition hohlraum plasmas, we find that the SBS will appear different behaviors. Due to strong Landau damping in H or HeH plasmas, the SBS is the strong convective instability, while the SBS in AuB plasmas is the absolute instability due to the weak Landau damping of the fast mode in AuB plasmas. However, although the SBS in CH plasmas is weak convective instability in linear theory, the large IAW excited by strong SBS will trap particles, which will flatten the particles distribution around the IAW phase velocity and decrease the linear Landau damping of IAW \cite{Neil_1965POF,Feng_2016POP}, therefore, the weak convective instability will transform into the absolute instability. The strong convective SBS can be controlled by increasing the linear Landau damping of IAW \cite{Neumayer_2008PRL,Neumayer_2008POP}. While the absolute SBS or the weak convective SBS should be controlled by the nonlinear saturation mechanisms, such as particles trapping, harmonics trapping, nonlinear frequency shift, wave breaking and so on.

This paper is organized as follows: In Sec. \ref{Sec:Theory analysis}, the kinetic theoretical analyses of IAW modes and SBS key feature parameters in different species plasmas are given. In Sec. \ref{Sec:Vlasov simulation}, the simulation by Vlasov-maxwell code is carried out to verify the kinetic theoretical analyses of IAW modes excited by SBS in different species plasmas. Then our simulations research the different SBS behaviors in different species ignition hohlraum plasmas in the high-temperature and high-density region. At last, the nonlinear behaviors in CH and AuB plasmas have been researched. In Sec. \ref{Sec:Summary}, the main conclusions are given.

\section{\label{Sec:Theory analysis}Theoretical analysis}
\subsection{\label{subsection:IAW}Modes analysis of ion-acoustic wave in stimulated Brillouin scattering}
In the stimulated Brillouin scattering (SBS) process, the pump light (EMW0) will resonantly decay into an IAW and an inverse scattering EMW (denoted as EMWs), the three waves satisfy the energy and momentum conservation, i.e., SBS:  $EMW0 [\vec{k}_{0}, \omega_{0}]\rightarrow EMWs [\vec{k}_{s}, \omega_{s}]+IAW [\vec{k}_{A}, \omega_{A}]$, where $[\vec{k}_i, \omega_i]$ are the wave vectors and the frequencies of the corresponding waves. The matching condition of the three waves in the SBS is:
\begin{equation}
\label{Eq:matching}
\vec{k}_{0}=\vec{k}_s+\vec{k}_{A},\quad \omega_{0}=\omega_s+\omega_{A}.
\end{equation}
Because $\omega_{A}\ll\omega_0$, and the direction of the wave vectors of the pump light and the scattered light are inverse for backward SBS discussed here, Eq. (\ref{Eq:matching}) can be written as:
\begin{equation}
\label{Eq:wavenumber}
k_s\simeq-k_{0}, \quad k_{A}\simeq 2k_{0}.
\end{equation}
From the dispersion relation of EMW0:
\begin{equation}
\omega_{0}^2
=\omega_{pe}^2+k_{0}^2c^2,
\end{equation}
one can obtain the wave number of the pump light $k_0$ in a given plasmas:
\begin{equation}
\label{Eq:k_0}
k_0\lambda_{De}=\frac{v_{te}}{c}\sqrt{n_c/n_e-1},
\end{equation}
where $v_{te}=\sqrt{T_e/m_e}$ is the electrons thermal velocity, $n_e, T_e, m_e$ is the density, temperature and mass of the electron. Thus, from Eqs. (\ref{Eq:wavenumber}) and (\ref{Eq:k_0}), one can obtain
\begin{equation}
\label{Eq:k_A}
k_A\lambda_{De}\simeq2\frac{v_{te}}{c}\sqrt{n_c/n_e-1}.
\end{equation}

\begin{figure}[!tp]
	\includegraphics[width=1\columnwidth]{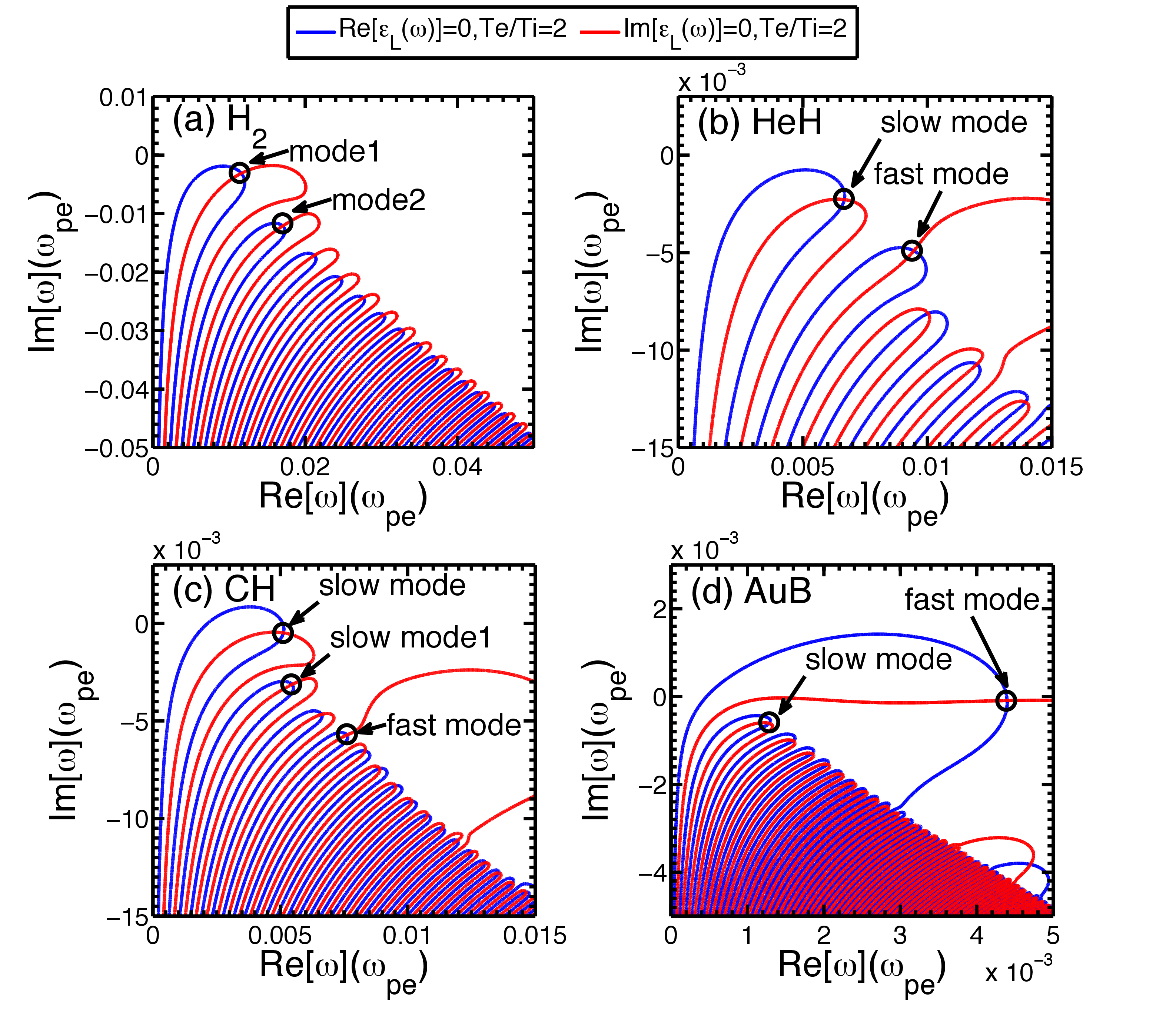}
	
	\caption{\label{Fig:Dispersion}(Color online) Contours of solutions to the fast and slow IAW mode dispersion relations of (a) H$_2$, (b) HeH, (c) CH and (d) AuB plasmas. Assuming that all the atoms are fully ionized. The conditions are: $T_e=5keV$, $T_i=0.5T_e$, $n_e=0.3n_c$, and $k_{A}\lambda_{De}=0.3$.}
\end{figure}
The linear kinetic theory of ion acoustic wave in a homogeneous, non-magnetized plasma consisting of multi-species ions is considered. Considering fully ionized, neutral, unmagnetized plasmas with the same temperature of all ion species ($T_H=T_C=T_i$), the linear frequencies and Landau damping of the plasmas are then given by the zeros of the plasma dielectric function, i.e., $\epsilon(k,\omega)=0$, which gives the linear dispersion relation of the ion acoustic wave in multi-ion species plasmas
\begin{equation}
\label{Eq:Dispersion}
\epsilon_L(\omega,k=k_A)=1+\chi_e+\sum_\beta \chi_{i\beta}=0,
\end{equation}
where $\chi_e$ is the susceptibility of electron and $\chi_{i\beta}$ is the susceptibility of ions $\beta$. $\chi_j$ (j present for electron or ion) can be expressed by $Z$ function
\begin{equation}
\label{Eq:Chi}
\chi_j=\frac{1}{(k\lambda_{Dj})^2}(1+\xi_jZ(\xi_j)),
\end{equation}
where a Maxwellian velocity distribution for all  species is assumed. $\xi_j=\omega/(\sqrt{2}kv_{tj})$ is generally complex, $\omega=Re(\omega)+iIm(\omega)$, $k$ is the frequency and the wave number of the given mode (such as the fast or slow IAW mode). $v_{tj}=\sqrt{T_j/m_j}$ ($T_j$, $m_j$ are the temperature and the mass of particle $j$) is the thermal velocity of particle $j$. $\lambda_{Dj}$ is the Debye length $\lambda_{Dj}=\sqrt{T_j/4\pi n_jZ_j^2e^2}$, i.e., $\lambda_{Dj}=v_{tj}/\omega_{pj}$ ($\omega_{pj}=\sqrt{4\pi n_jZ_j^2e^2/m_j}$ is the plasma frequency for specie j). $Z$ function is the dispersion function 
\begin{equation}
\label{Eq:Zeta}
\begin{aligned}
Z(\xi_j)  =\frac{1}{\sqrt{\pi}}\int_{-\infty}^{+\infty}\frac{e^{-v^2}}{v-\xi_j}dv
 =i\sqrt{\pi}e^{-\xi_j^2}(1+erf(i\xi_j)),
\end{aligned}
\end{equation}
in which $erf(i\xi_j)=2/\sqrt{\pi}\int_0^{i\xi_j} e^{-\eta^2}d\eta$ is the error function, $\xi_j$ is a complex variable. A newly developed accurate algorithm showed by M. R. Zaghloul \cite{Zaghloul_2012ACM} is taken to calculate the Faddeyeva (plasma dispersion) function $\omega(z)=e^{-z^2}(1+erf(iz))$ (where $z=x+iy$ is a complex variable).

Under the condition of $T_e=5keV$, $n_e=0.3n_c$, one can obtain the wave number of the IAW $k_A\lambda_{De}=0.3$ from Eq. (\ref{Eq:k_A}). By solving Eq. (\ref{Eq:Dispersion}), the contours of solutions of the IAWs in four typical plasmas are shown in Fig. \ref{Fig:Dispersion} under the condition of $T_i=0.5T_e$, $k_A\lambda_{De}=0.3$. Although there are infinite solutions of the IAW as shown in Fig. \ref{Fig:Dispersion}, the mode with the least Landau damping ($|Im(\omega)|$) will be preferentially excited in SBS. There exists only one group of modes in H plasmas as shown in Fig. \ref{Fig:Dispersion}(a), while there exist two groups of modes called \textquotedblleft the fast mode" and \textquotedblleft the slow mode" in multi-ion species plasmas as shown in Figs. \ref{Fig:Dispersion}(b)-\ref{Fig:Dispersion}(d). Here, the fast mode and the slow mode refer to the least damped modes belonging to each class of mode. In H plasmas, the least damped mode is called \textquotedblleft mode 1" because there is no fast mode or slow mode branches. The least damped mode in HeH or CH plasmas under the condition of $T_i/T_e=0.5$ is the slow mode, while the least damped mode in AuB plasmas is the fast mode. The parameters of the least damped modes in different species plasmas are listed in Table \ref{table1}. 
\begin{table*}
	\caption{\label{table1} The IAW parameters and convection or absolute instability parameters of SBS in different plasmas. Assuming that all the atoms are fully ionized. The conditions are: $T_e=5keV$, $T_i=0.5T_e$, $n_e=0.3n_c$, $k_{A}\lambda_{De}=0.3$, and $I_0=3\times10^{15}W/cm^2$. }
	
	\begin{ruledtabular}
		
		\begin{tabular}{ccccc|ccc|c}
			\hline
			\multirow{3}*{Plasmas}& \multicolumn{7}{|c|}{Theory} & \multirow{2}*{Simulation}\\
			\cline{2-8}
			&\multicolumn{4}{|c|}{IAW\footnotemark[1]}&\multicolumn{3}{c|}{SBS}
			\\
			\cline{2-9}

			& \multicolumn{1}{|c}{$Re(\omega)/(10^{-3}\omega_{pe})$} &$Re(\omega)/(10^{-3}\omega_{0})$& $Im(\omega)/(10^{-3}\omega_{pe})$ &$Im(\omega)/Re(\omega)$ \footnotemark[2]& $\omega_s/\omega_0$& $\nu_A/(2\gamma_B)$\footnotemark[3]& $G_B$& $\omega_s/\omega_0$
			\\
			\hline
			
			\multicolumn{1}{c}{$\text{H}_2$\footnotemark[4]}& \multicolumn{1}{|c}{11.6} & 6.35 & -3.13 & -0.270 & 0.99365 & 6.99 & 3.67 & 0.9936
			\\
			
			\multicolumn{1}{c}{HeH}& \multicolumn{1}{|c}{6.67} & 3.65 & -2.28 & -0.342 & 0.99635 & 8.85 & 2.90 & 0.9963
			\\
			
			\multicolumn{1}{c}{CH}& \multicolumn{1}{|c}{5.12} & 2.80 & -0.49 & -0.096 & 0.99720 & 2.48 & 10.37 & 0.9972
			\\
			
			\multicolumn{1}{c}{AuB}& \multicolumn{1}{|c}{4.40} & 2.41 & -0.09& -0.020 & 0.99759 & 0.53 & 48.50 & 0.9975
			\\
			\hline
		\end{tabular}
		
		\footnotetext[1]{The IAW refers to the least damped mode.}
		
		\footnotetext[2]{$Re(\omega)\equiv\omega_A, |Im(\omega)|\equiv\nu_A$.}
		\footnotetext[3]{$\nu_A/(2\gamma_B): 1/2*\nu_A/\gamma_{0B}*(\sqrt{v_{gB}/v_{gA}})$.}
		\footnotetext[4]{$\text{H}_2$ refers to H plasmas.}
	\end{ruledtabular}
	
\end{table*}

\subsection{\label{subsection:SBS}Absolute and convective condition for stimulated Brillouin scattering}
Assuming the plasmas are local uniform, the standard linear one-dimensional SBS three-waves interaction equations in homogenous plasmas are \cite{Berger_1998POP,HaoLiang_2012CSB,Forslund_1975POF}
\begin{equation}
\label{Eq:A0}
(\frac{\partial}{\partial t}+v_{g0}\frac{\partial}{\partial x}+\nu_0)A_0(x, t)=-i\frac{\pi e^2}{\omega_0m_e}\delta n_AA_s,
\end{equation}
\begin{equation}
\label{Eq:As}
(\frac{\partial}{\partial t}-v_{gs}\frac{\partial}{\partial x}+\nu_s)A_s(x, t)=-i\frac{\pi e^2}{\omega_sm_e}\delta n_A^*A_0,
\end{equation}
\begin{equation}
\label{Eq:nA}
(\frac{\partial}{\partial t}+v_{gA}\frac{\partial}{\partial x}+\nu_A)\delta n_A(x, t)=-i\frac{\bar{Z}_in_ee^2k_A^2}{4\omega_Am_e\bar{m}_ic^2}A_0A_s^*,
\end{equation}
where $A_0(x,t)$, $A_s(x,t)$ and $\delta n_A(x,t)$ are the complex amplitudes of the vector potentials for the pump light, SBS backscattering light, and perturbation of the electron density $n_e$ with the frequency $\omega_A$ from IAW. And $v_{gi}$, $\nu_i$, $\omega_i$ are the group velocities, damping rates and frequencies of the pump light ($i=0$), SBS scattering light ($i=s$), and IAW ($i=A$). Assuming that the ions average mass and average charge number are $\bar{m}_i$ and $\bar{Z}_i$.
Following Hao et al. \cite{HaoLiang_2012CSB,Hao_2013LPB}, one can obtain the usual SBS threshold $\gamma_{thB}=\sqrt{\nu_A\nu_s}$ and the threshold of absolute SBS instability is
\begin{equation}
\gamma_{aB}=\sqrt{v_{gs}v_{gA}/4}(\nu_s/v_{gs}+\nu_A/v_{gA})\simeq\frac{1}{2}\nu_A\sqrt{\frac{v_{gs}}{v_{gA}}},
\end{equation}
in general, $\nu_s/v_{gs}\ll\nu_A/v_{gA}$, thus $\nu_s/v_{gs}$ can be neglected. The condition for convective SBS is \cite{Forslund_1975POF}
\begin{equation}
\gamma_{thB}<\gamma_{0B}<\gamma_{aB},
\end{equation}
where
\begin{equation}
\gamma_{0B}=\frac{1}{4}\sqrt{\frac{n_e}{n_c}}\frac{v_0}{v_{te}}\sqrt{\omega_0\omega_A}
\end{equation}  
is the maximum temporal growth rate of SBS \cite{Berger_2015PRE,Lindl_2004POP},  $v_0=eA_0/m_ec$ is the electrons quiver velocity. Thus, the parameter $\nu_A/(2\gamma_B)\equiv1/2*\nu_A/\gamma_{0B}*\sqrt{v_{gs}/v_{gA}}$ can be taken as an important factor to decide convective SBS or absolute SBS. If $\nu_A/(2\gamma_B)\gg1$, the SBS is the strong convective instability, such as SBS in $\text{H}_2$ or HeH plasmas in our simulation. If $\nu_A/(2\gamma_B)$ is not much larger than 1, the SBS is the weak convective instability, such as SBS in CH plasmas in our simulation. If $\nu_A/(2\gamma_B)<1$, the SBS is the absolute instability, such as SBS in AuB plasmas in our simulation. The gain of SBS is then given by \cite{Berger_2015PRE,Lindl_2004POP}
\begin{equation}
\label{Eq:G_B}
G_B=2\frac{\gamma_{0B}^2}{\nu_Av_{gs}}L=\frac{1}{8}\frac{v_0^2}{v_{te}^2}\frac{n_e}{n_c}\frac{\omega_A}{\nu_A}\frac{\omega_0}{v_{gs}}L,
\end{equation}
where $v_{gs}=c^2k_s/\omega_s$ is the group velocity of SBS scattering light, $L$ is the plasmas density scale length, $\omega_A\equiv Re(\omega)$ and $\nu_A\equiv |Im(\omega)|$ are the frequency and Landau damping of IAW, which are listed in Table \ref{table1}. The collision damping of IAW can be neglected since the electrons temperature is as high as $T_e=5keV$ in our simulation, thus only the Landau damping of IAW is considered.

 Because $G_B\propto\omega_A/\nu_A\equiv|Re(\omega)/Im(\omega)|$, in the same specie plasmas, the IAW mode with the least Landau damping $|Im(\omega)/Re(\omega)|$ will produce the largest gain and will increase most quickly in SBS. Thus, in general, the least damped mode will be excited in SBS. In HeH, H, CH and AuB plasmas, with the Landau damping $|Im(\omega)/Re(\omega)|$ of IAW decreasing, the gain of SBS will increase obviously. Especially, the SBS in AuB plasmas is absolute instability with the gain $G_B=48.5$ since the Landau damping of the fast mode in AuB plasmas is very weak. Under the strong damping condition, such as H or HeH plasmas, the SBS is strong convective instability with the gain $G_B\sim2-4$. Under the strong damping condition $\nu_A/\gamma_{0B}*\sqrt{v_{gs}/v_{gA}}\gg1$ \cite{Forslund_1975POF}, one can get the Tang model \cite{Tang_1966JAP} from Eqs. (\ref{Eq:A0})-(\ref{Eq:nA}):
 \begin{equation}
 R(1-R)=\epsilon\{exp[G_B(1-R)]-R\},
 \end{equation}
 where $R=I_s(x=0)/I_0(x=0)$ is the reflectivity of SBS at the left boundary (incident boundary), and $\epsilon=I_s(x=L)/I_0(x=0)$ is the boundary condition of the backscattering light wave at the right boundary (transmitting boundary), which is called seed light. If $R\ll1$, the Tang model can be approximate to the seed amplification equation:
 \begin{equation}
 \label{Eq:seed amplification}
 I_s(x=0)/I_s(x=L)=exp(G_B),
 \end{equation}
 where $I_s(x=0)$, $I_s(x=L)$ are the intensity of the SBS scattering light at the left boundary and the seed light at the right boundary. Below, the intensity of the incident light $I_0(x=0)$ and the seed light $I_s(x=L)$ are denoted as $I_0$ and $I_s$.

\section{\label{Sec:Vlasov simulation}Numerical simulation}
An one dimension in space and one dimension in velocity (1D1V) Vlasov-Maxwell code \cite{Liu_2009POP,Liu_2011POP,Liu_2012PPCF} is used to research the SBS behavior in typical ignition hohlraum plasmas. We choose the high-temrature and high-density region as an example, the electrons temperature and electrons density are $T_e=5keV$, $n_e=0.3n_c$, where $n_c$ is the critical density for the incident laser \cite{Liu_2011POP}. The electrons density is taken to be higher than $0.25n_c$, thus the stimulated Raman scattering and two-plasmon decay instability \cite{Xiao_2015POP,Xiao_2016POP} are excluded. The H, HeH, CH and AuB plasmas are taken as typical examples for they are common in ICF \cite{He_2016POP,Glenzer_2007Nature}. The ions temperature is $T_i=0.5T_e$. The linearly polarized laser intensity is $I_0=3\times10^{15}W/cm^2$ with wavelength $\lambda_0=0.351\mu m$. The spatial scale is [0, $L_x$] discretized with $N_x=5000$ spatial grid points and spatial step $dx=0.2c/\omega_0$. And the spatial length is $L_x=1000c/\omega_0\simeq160\lambda_0$ with $2\times5\%L_x$ vacuum layers and $2\times5\%L_x$ collision layers in the two sides of plasmas boundaries. The plasmas located at the center with density scale length $L=0.8L_x$ are collisionless. The incident laser propagates along the $x$ axis from the left to the right with outgoing boundary conditions. The strong collision damping layers are added into the two sides of the plasmas boundaries ($2\times5\%L_x$) to damp the electrostatic waves such as IAWs at the boundaries and decrease the effect of sheath field. The electrons velocity scale $[-0.8c, 0.8c]$ and the ions velocity scale $[-0.03c, 0.03c]$ are discretized with $2N_v+1$ ($N_v=512$) grid points. The total simulation time is $t_{end}=6\times10^4\omega_0^{-1}$ discretized with $N_t=3\times10^5$ and time step $dt=0.2\omega_0^{-1}$.
  \begin{figure}[!tp]
  	\includegraphics[width=1\columnwidth]{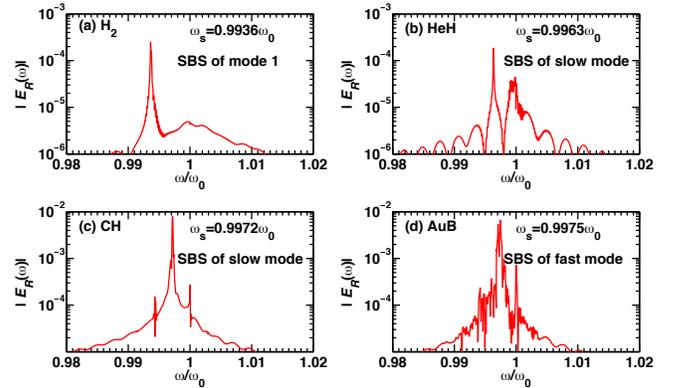}
  	
  	\caption{\label{Fig:Spectra_w}(Color online) The frequency spectra of scattering light in four cases: (a) H$_2$, (b) HeH, (c) CH, and (d) AuB plasmas.
  		The parameters are $n_e=0.3n_c, T_e=5keV, I_0=3\times10^{15}W/cm^2$ as the same as Fig. \ref{Fig:Dispersion}. The seed light intensity is $I_s=1\times10^{-4}I_0$ and with the matched frequency calculated by theory in Table \ref{table1}.}
  \end{figure}
  
 Figure \ref{Fig:Spectra_w} shows the spectra of the SBS scattering lights in the cases of H, HeH, CH, and AuB plasmas. The peak frequecy of scattering light in H plasmas is $\omega_s=0.9936\omega_0$, which is consistent to the theoretical frequency of SBS of mode 1 in H plasmas as shown in Table \ref{table1}. In HeH, CH, AuB plasmas, the peak frequencies of SBS scattering lights are $\omega_s/\omega_0=0.9963, 0.9972, 0.9975$ respectively, which are consistent to the theoretical values $\omega_s/\omega_0=0.99635, 0.99720, 0.99759$. However, the least damped mode in HeH or CH plasmas is the slow mode in our simulation, while the least damped mode in AuB plasmas is the fast mode. These simulations verify that the theoretical IAW frequencies calculated in part \ref{subsection:IAW} are accurate both for the single-ion species and multi-ion species plasmas. The least damped IAW mode will get the largest gain as shown in Eq. (\ref{Eq:G_B}), thus the SBS in HeH or CH plasmas is from the slow mode, while the SBS in AuB plasmas is from the fast mode. Although the seed light with the least damped mode frequecy is taken to promote  SBS here, we have also tested the cases without seed light (not shown here), no matter whether the seed light is added or not, the conclusions are not affected except that the spectra amplitudes are different especially in H or HeH plasmas.

  \begin{figure}[!tp]
  	\includegraphics[width=1\columnwidth]{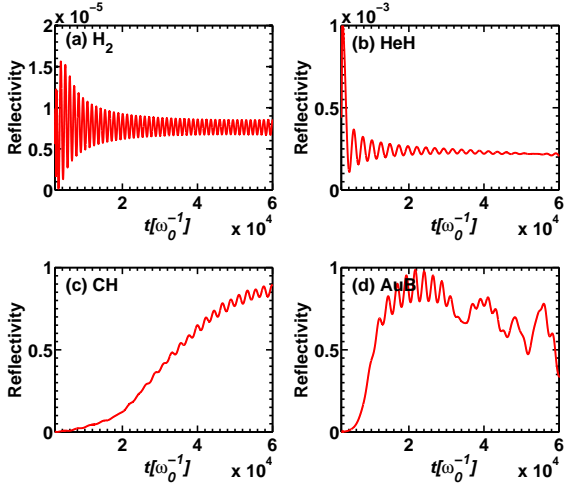}
  	
  	\caption{\label{Fig:Reflectivity}(Color online) The SBS reflectivity in the frequency scope $[0.9\omega_0, 0.999\omega_0]$ for the four plasmas: (a) H$_2$, (b) HeH, (c) CH, and (d) AuB plasmas. The weak reflected light with frequency of $\omega_0$ is filtered from the backward scattering lights. In case (a) H$_2$, the seed light intensity is $I_s=1\times10^{-6}I_0$, while (b)-(d): $I_s=1\times10^{-4}I_0$.}
  \end{figure}
\begin{figure}[!tp]
  	\includegraphics[width=1\columnwidth]{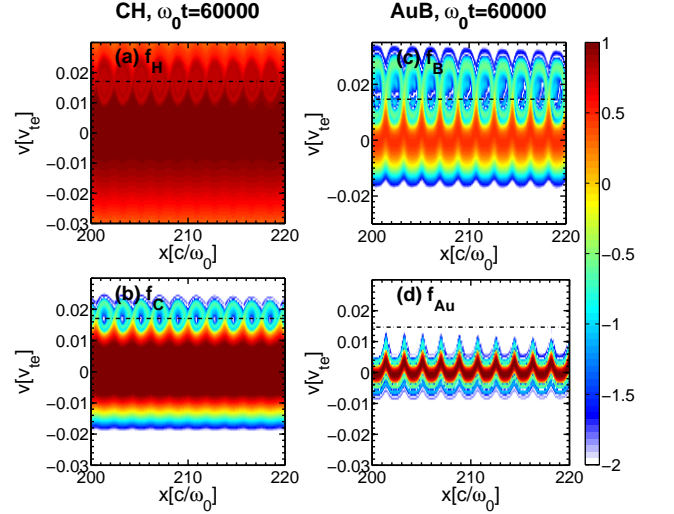}
  	
  	\caption{\label{Fig:phas pictures}(Color online) The distributions of (a) H ions, (b) C ions in CH plasmas and (c) B ions, (d) Au ions in AuB plasmas at the time of $\omega_0t=60000$.
  		The black dashdotted lines are the phase velocities of slow IAW mode for CH plasmas in (a) (b), and fast IAW mode for AuB plasmas in (c) (d).}
  \end{figure}

 Figure \ref{Fig:Reflectivity} gives a snapshot of the SBS reflectivity in the typical four plasmas. For H or HeH plasmas, $\nu_A/2\gamma_B\gg1$, the SBS is strong convective instability, and the gain of SBS is as low as $2-4$ as shown in Table \ref{table1}. Thus, the SBS reflectivity in H or HeH plasmas is very low if the seed light intensity $\epsilon=I_s/I_0=1\times10^{-6}$ or $1\times10^{-4}$. Because $R\ll1$ in the cases of H or HeH plasmas, the gain of SBS can be calculated from Eq. (\ref{Eq:seed amplification}). Taking H plasmas as an example, the gain is $G_B\simeq ln(R/\epsilon)\simeq ln(8\times10^{-6}/(1\times10^{-6}))=2.08$, which is slightly lower than the theoretical value $G_B=3.67$. Maybe the reason is that the resonance length in the simulation is shorter than the plasmas density scale length. For CH plasmas, $\nu_A/(2\gamma_B)=2.48$ is not much larger than 1, the SBS is weak convective instability. Because the gain of SBS $G_B=10.37$ is large, the SBS will excite strong IAW. The strong IAW will trap ions or electrons as shown in Fig. \ref{Fig:phas pictures}, thus the linear Landau damping of IAW will be decreased due to particles flatting around the IAW phase velocity \cite{Feng_2016POP,Neil_1965POF}. The convective parameter $\nu_A/(2\gamma_B)$ will decrease below 1,  the weak convective instability will transform into absolute instability. However, for AuB plasmas, $\nu_A/(2\gamma_B)=0.53<1$, the SBS is absolute instability, and the gain of SBS $G_B=48.5$ is very large, thus the SBS will increase quickly and saturate at the later time as shown in Fig. \ref{Fig:Reflectivity}(d). The nonlinear saturation mechanisms mainly include particles trapping, harmonics generation and nonlinear frequency shift, which will be discussed in detail below.
 
Figure \ref{Fig:phas pictures} demonstrates the phase pictures of ions in CH and AuB plasmas. Due to ions trapping, it will produce two effects: (1) the linear Landau damping will be decreased \cite{Feng_2016POP}, which will increase the SBS gain; (2) the IAW frequency will shift \cite{Feng_2016PRE}, which will decrease the SBS by detuning the SBS three-waves matching. Thus the trapping effect will first increase the SBS by decreasing the IAW Landau damping and then saturate the SBS by nonlinear frequency shift (NFS). The trapping effects on SBS have been researched in detail by Williams et al. \cite{Williams_2004POP}. Since the least damped mode in CH plasmas is the slow mode while the least damped mode in AuB plasmas is the fast mode, the slow IAW mode in CH plasmas and the fast IAW mode in AuB plasmas will be excited through SBS. The black dashdotted lines represent the phase velocities of the slow mode in CH plasmas and the fast mode in AuB plasmas. We can see that the ions will be trapped around the  IAW phase velocity. Due to large IAW amplitude in AuB plasmas, the trapping effect will lead to an obvious nonlinear frequency shift as shown in Fig. \ref{Fig:phas pictures}(c). Since the fast mode phase velocity $v_{\phi}=Re(\omega)/k_A$ is nearly 12 times the Au ions thermal velocity $v_{tAu}=\sqrt{T_{Au}/m_{Au}}$, the number of Au ions around the fast mode phase velocity is very small thus Au ions can nearly not be trapped by the fast IAW mode. However, the slow IAW phase velocity in CH plasmas is nearly 3.6 times the C ions thermal velocity and nearly equals to the H ions thermal velocity, thus the slow IAW mode can trap both the C ions and H ions in CH plasmas.
\begin{figure}[!tp]
	\includegraphics[width=1\columnwidth]{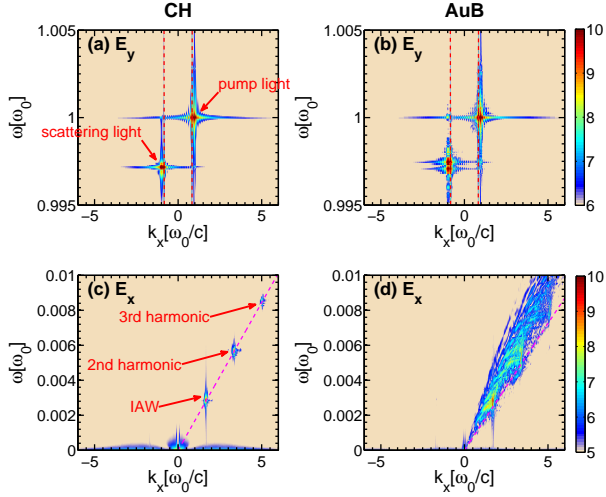}
	
	\caption{\label{Fig:w_k}(Color online) The dispersion relation of (a), (b) transverse electric field $E_y$ of light and (c), (d) longitudinal electrostatic field $E_x$ of IAW. (a), (c): CH plasmas; (b), (d): AuB plasmas. The red dashed lines are the EMW dispersion relations in (a) and (b), and the pink dashed lines are the IAW dispersion relations in (c) and (d).}
\end{figure}

In addition to the particles trapping, the harmonics generation of IAW is also an important saturation mechanism of the SBS. As shown in Fig. \ref{Fig:w_k}, only the fundamental IAW mode will resonantly couple with the pump light to produce the scattering light, while the harmonics such as the second harmonic and the third harmonic of IAW do not satisfy the three-waves matching condition. The fundamental IAW mode energy will transfer a part to the harmonics so that the fundamental IAW mode amplitude and SBS will be decreased due to harmonics generation. On the other hand, the harmonics generation will also lead to the fluid nonlinear frequency shift of nonlinear IAW \cite{Feng_2016PRE, Berger_2013POP,Chapman_2013PRL}, thus the fluid NFS due to harmonics generation will detune the SBS and saturate the SBS. In our simulation, $k_A\lambda_{De}=0.3$ is in the kinetic region, thus the kinetic NFS due to particles trapping will dominate in the total NFS \cite{Feng_2016PRE} and determine the saturation of SBS. For AuB plasmas, since the Landau damping of the IAW is very low and the gain of SBS is very high, the SBS is very strong and produce very strong IAW. The very strong IAW will produce strong harmonics and large trapping width ($\Delta v_{tr}=2\sqrt{q_i\phi/m_i}$, $\phi$ is the IAW electric potential) \cite{Berger_2013POP}. Thus the positive NFS of IAW mainly from particles trapping is obvious as shown in Fig \ref{Fig:w_k}(d), so that the scattering light frequency of SBS will appear a negative shift as shown in Fig. \ref{Fig:w_k}(b).

   \begin{figure}

   	   	\includegraphics[width=1\columnwidth]{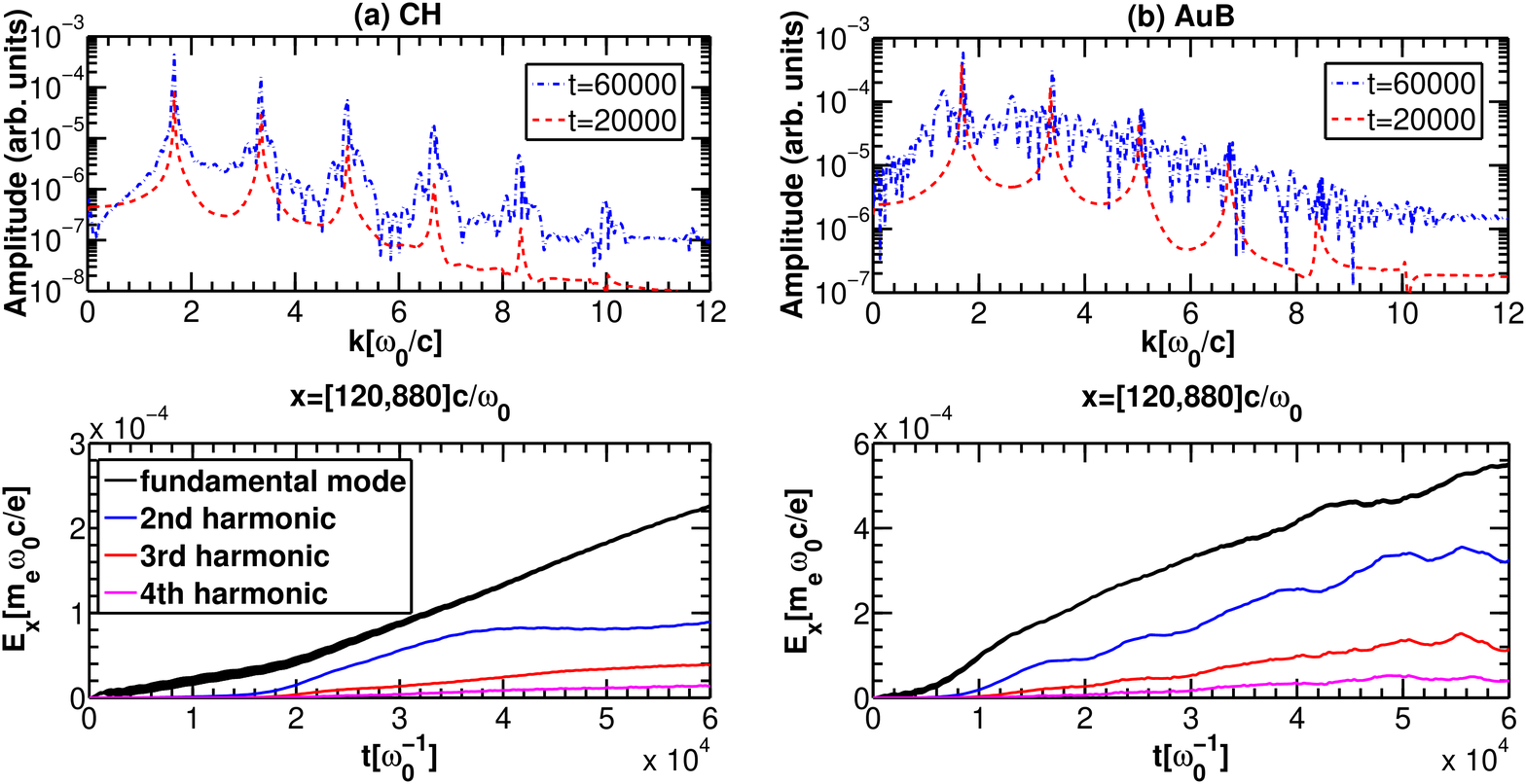}

   	\caption{\label{Fig:spectral_k}(Color online) The wave number spectra (upper panel) and corresponding harmonics evolution with time (lower panel) of electrostatic field $E_x$ for (a) CH plasmas and (b) AuB plasmas. Where the wave number scales of the fundamental mode, the first, second, third harmonics are $k_1\in[0.8, 2.4]\omega_0/c$, $k_2\in[2.4, 4.0]\omega_0/c$, $k_3\in[4.0, 5.6]\omega_0/c$, $k_4\in[5.6, 7.2]\omega_0/c$ averaged in the spatial scale $x=[120, 880]c/\omega_0$.}
   \end{figure}
   
%

The wave number spectra and harmonics evolution with time are demonstrated in Fig. \ref{Fig:spectral_k}. In the early stage of SBS, such as $t=20000\omega_0^{-1}$, the harmonics are discrete as shown in the wave number spectra of $E_x$. However, in the late stage of SBS, such as $t=60000\omega_0^{-1}$, the harmonics in CH plasmas keeps discrete as shown in the upper panel of Fig. \ref{Fig:spectral_k}(a), while the harmonics in AuB plasmas are widened and appear continuous spectra as shown in the upper panel of Fig. \ref{Fig:spectral_k}(b), which is consistent to the spectra in Figs. \ref{Fig:w_k}(c) and \ref{Fig:w_k}(d).
Because the Landau damping of the fast mode in AuB plasmas is much lower than the slow mode in CH plasmas as shown in Table \ref{table1}, the IAW amplitude in AuB plasmas develops quicker than in CH plasmas as shown in the lower panels of Figs. \ref{Fig:spectral_k}. The turbulence would be easier to occur in the large IAW amplitude in AuB plasmas, which leads to the widening of IAW spectra.  Although the fundamental mode and harmonics in AuB plasmas are stronger than that in CH plasmas, the spectra broadening in AuB plasmas will lead to larger energy dissipation of the fundamental IAW mode. Thus the harmonics energy dissipation and spectra broadening are also important factors to saturate SBS especially for the absolute instability.

\begin{figure}[!tp]
 	\includegraphics[width=1\columnwidth]{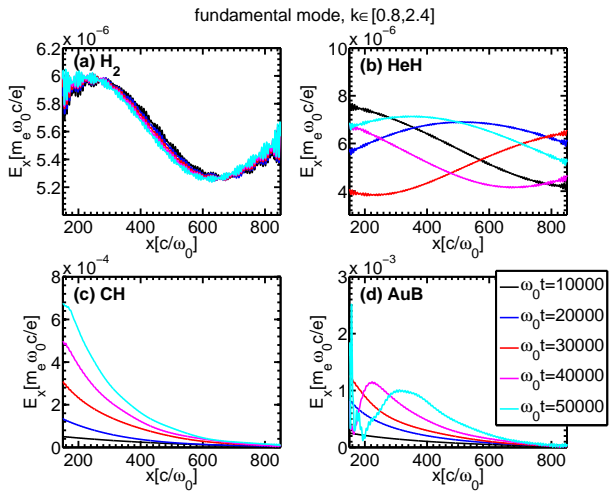}
 	
 	\caption{\label{Fig:Ex_x}(Color online) The spatial distribution of fundamental IAW mode electric field at different time for (a) H$_2$, (b) HeH, (c) CH and (d) AuB plasmas. The parameters are the same as Fig. \ref{Fig:Reflectivity}.}
 \end{figure}

To verify that the SBS in H or HeH plasmas is indeed convective instability and the SBS in CH or AuB plasmas is absolute instability, the spatial evolution of the fundamental IAW mode is shown in Fig. \ref{Fig:Ex_x}. Since the SBS in H or HeH plasmas satisfies strong damping condition $\nu_A/\gamma_B\gg1$, the SBS in H or HeH plasmas is strong convective instability. As shown in Figs. \ref{Fig:Ex_x}(a) and \ref{Fig:Ex_x}(b), the fundamental IAW mode appears the feature of convective instability.  Due to the continuous seed light, the fundamental IAW mode oscillates in the given space region with the time. For CH plasmas, since $\nu_A/\gamma_B$ is not much larger than 1 and the SBS gain $G_B=10.37$ is large, the SBS is weak convective instability in linear theory. However, due to particles trapping by IAW generated by strong SBS (as shown in Figs. \ref{Fig:phas pictures}(a) and \ref{Fig:phas pictures}(b)), the linear Landau damping of IAW will be decreased \cite{Neil_1965POF,Feng_2016POP} after several ion bounce time $\tau_{bi}=2\pi/\sqrt{k_Aq_iE_x/m_i}$. When the Landau damping of IAW in CH plasmas decreases to the level that the absolute instability condition is satisfied, i.e., $\nu_A/(2\gamma_B)<1$, the SBS will transform to absolute instability. In our simulation, there indeed exists obvious particles trapping around the IAW in CH plasmas as shown in Figs. \ref{Fig:phas pictures}(a) and \ref{Fig:phas pictures}(b), the IAW appears the features of absolute instability. That is to say, in every space position, the IAW amplitude will increase with time as shown in Fig. \ref{Fig:Ex_x}(c). For AuB plasmas, the initial condition $\nu_A/(2\gamma_B)<1$, thus the SBS in AuB plasmas is absolute instability and the SBS gain $G_B=48.5$ is very large. As a result, the IAW amplitude in AuB plasmas increases more quickly than that in CH plasmas as shown in Figs. \ref{Fig:Ex_x}(c) and \ref{Fig:Ex_x}(d). When $t\gtrsim30000\omega_0^{-1}$, the amplitude of the IAW near the left boundary reaches as large as $E_x\simeq1\times10^{-3}m_e\omega_0c/e$, maybe the wave breaking limit is reached so that the IAW amplitude can not be increased further, as a result, the local IAW amplitude near the left boundary decreases rapidly. That is also a main reason for the SBS saturation in the late stage of SBS. However, before reaching the wave breaking limit, the IAW appears obvious features of absolute instability as shown in Fig. \ref{Fig:Ex_x}(d). For example, before $t\simeq30000\omega_0^{-1}$, in every space position the IAW amplitude will increase with time. And except for the wave breaking region near the left boundary, for example, in the space region $x\in[300, 800]c/\omega_0$, the IAW amplitude will also increase with time.

\section{\label{Sec:Summary}Summary}
In conclusions, there exists only one group of modes in H plasmas, while there exist two groups of modes in multi-ion species plasmas. In our simulation, the least damped mode in HeH or CH plasmas is the slow mode, while the least damped mode in AuB plasmas is the fast mode, the least damped mode will be excited preferentially in SBS. Different instability features of SBS in the high-temperature and high-density region in H (or HeH) plasmas, CH plasmas, AuB plasmas have been analyzed. In H or HeH plasmas, the SBS is strong convective instability due to the strong Landau damping. In AuB plasmas, the SBS is absolute instability due to the weak Landau damping. While in CH plasmas, although the SBS is weak convective instability, since the linear Landau damping will be decreased by particles trapping, the SBS in CH will transform to absolute instability and the SBS behavior appears the absolute instability features. Several nonlinear saturation mechanisms such as particles trapping, harmonics generation and nonlinear frequency shift are also demonstrated to explain the saturation of absolute SBS. These results give a guidance to the IAW modes excited by SBS and different SBS behaviors in different species plasmas especially in the high-temperature and high-density region and provide the possibility to control SBS by changing ion species.

\begin{acknowledgments}
We are pleased to acknowledge useful discussions with K. Q. Pan, L. Hao, J. X. Gong and B. Qiao. This research was supported by the National Natural Science Foundation of China (Grant Nos. 11375032, 11575035, 11475030 and 11435011), National Basic Research
Program of China (Grant No. 2013CB834101) and science challenge project, No. JCKY2016212A505. 
\end{acknowledgments}

\bibliography{SBS_multi_species}

\end{document}